\documentclass[fleqn,usenatbib]{mnras}

\usepackage{newtxtext,newtxmath}

\usepackage[T1]{fontenc}

\DeclareRobustCommand{\VAN}[3]{#2}
\let\VANthebibliography\thebibliography
\def\thebibliography{\DeclareRobustCommand{\VAN}[3]{##3}\VANthebibliography}

\usepackage{graphicx}	

\newcommand{\angstrom}{\mbox{\normalfont\AA}}

\title[EK~Cep and HS~Her]{Eccentric orbits and apsidal motion in the eclipsing binaries \\
EK~Cep and HS~Her}

\author[Latkovi{\' c} et al.]{
Olivera Latkovi{\' c},$^{1}$\thanks{E-mail: olivia@aob.rs}
Kosmas Gazeas,$^{2}$
Haralambi Markov,$^{3}$
Atila {\v C}eki$^{1}$
and Sofia Palafouta$^{2}$
\\
$^{1}$Astronomical Observatory, Volgina 7, 11060 Belgrade, Serbia\\
$^{2}$Section of Astrophysics, Astronomy and Mechanics, Department of Physics, National and Kapodistrian University of Athens, GR-15784 Zografos, Athens, Greece\\
$^{3}$Institute of Astronomy and National Astronomical Observatory, Bulgarian Academy of Sciences, BG-1784 Sofia, Bulgaria}

\date{Accepted XXX. Received YYY; in original form ZZZ}

\pubyear{2015}

\begin{document}
\label{firstpage}
\pagerange{\pageref{firstpage}--\pageref{lastpage}}
\maketitle

\begin{abstract}
We present the first modern analysis of two young, eclipsing binaries, EK~Cep and HS~Her, based on new, ground-based, CCD multicolour light curves as well as the TESS observations, radial velocity curves, and eclipse timing measurements. The orbital and stellar parameters of the stars are determined by Roche modelling, and their evolutionary status is examined using a grid of isochrones and evolutionary tracks. We find that HS~Her is 25-32 Myr old and its components are on the zero-age main sequence; at the age of 16-20 Myr, the primary of EK~Cep is also on the ZAMS, but its secondary is a pre-main-sequence star. Both binaries have slightly eccentric orbits and display apsidal motion. Based on updated eclipse timings and spectroscopic evidence, we rule out the presence of a previously hypothesized tertiary component in HS~Her.
\end{abstract}


\section{Introduction}

EK~Cep (HD 206821, RA: 21$^h$ 41$^m$ 21$^s$.50, Dec: +69$^{\circ}$ 41$^{'}$ 34$^{''}$.11, P = 4.427794 d, $V_{mag}$ = 7.89) and HS~Her (HD 174714, RA: 18$^h$ 50$^m$ 49$^s$.77, Dec: +24$^{\circ}$ 43$^{'}$ 11$^{''}$.94, P = 1.6374352 d, $V_{mag}$ = 8.56) are detached eclipsing binaries in early stages of evolution. Both objects have interesting peculiarities. EK~Cep was the first known eclipsing binary with a pre-main-sequence (PMS) component, apsidal motion and metallicity measurements \cite[see e.g.,][]{claret1995}. HS~Her was the subject of a long debate surrounding its relatively short apsidal motion period (of the order of 100 years) and the possibility of having a third component \citep{colak2005}; there are also indications of primordial protostellar matter lingering inside the Roche lobe of the secondary star \citep{cakirli2007}.

These systems were examined in many previous works, mostly based on photometric light curves of relatively poor quality, obtained before the CCD era. The majority of the early studies extract their results from data with high noise and sparse phase coverage that was, in some cases, insufficient to detect the secondary eclipse. Most of these works rely on effectively deprecated modeling methods that preceded the widespread use of the Roche formalism. Early spectroscopic studies with radial velocity measurements are reported without uncertainties, which results in overestimated and unrealistic accuracy for the derived stellar parameters. An important weakness of the older studies is that the secondary component of HS~Her was not yet spectroscopically detected at the time when we initiated our own spectroscopic observations (in 2002; it was found for the first time in the study by \citealt{cakirli2007}). Presently, both systems investigated in this work have been neglected for more than a decade.

We conduct the first modern study of EK~Cep and HS~Her, based on the simultaneous Roche modelling of new CCD light curves, including TESS observations, and new radial velocity measurements, supported by a detailed eclipse timing analysis of all the measurements available in the literature and over 60 new eclipse timing measurements, contributed as part of this work. In the following sections, we provide an overview of the past results (sections \ref{secEKCepLit} and  \ref{secHSHerLit}), the description of the observations (sections \ref{secPhoto} and \ref{secSpectro}), the methods employed in our analyses (sections \ref{secET} and \ref{secModeling}), and proceed to discuss the evolutionary status (section \ref{secEvolution}) of the two systems under study.

\subsection{EK~Cep}
\label{secEKCepLit}

The first mention of EK~Cep is found in the work of \citet{ebbi1966a}, who published photomultiplier light-curves in the $B$ and $V$ passbands, determined the orbital period of about 4.4 days and noted the eccentricity of the system.
A spectroscopic study followed soon after \citep{ebbi1966b}, resulting in a single-lined radial velocity curve. The eccentricity was confirmed to be around 0.1. \citet{hill1975} classified EK~Cep as A1 V star.
\citet{mezzetti1980} reanalysed the light and radial velocity curves from \citet{ebbi1966a,ebbi1966b} with Wood's triaxial ellipsoids model \citep{wood1973} and remarked on the disagreement between the ratio of component radii with the expectations if both stars are on the main sequence (A0 or A1 and G1).

\citet{tomkin1983} was the first to detect the secondary component in the spectra and derive a double-lined radial velocity curve. Adopting the photometric solution of \citet{mezzetti1980}, they arrived at the following absolute parameters: $M_1=2.03 M_{\odot}$, $M_2=1.12 M_{\odot}$, $R_1=1.31 R_{\odot}$, $R_2=1.08 R_{\odot}$ and $a=16.63 R_{\odot}$. The secondary was determined to be roughly between F5 V and G5 V.
\citet{hill1984} conducted new photomultiplier observations in V and R filters and analysed them with the \textsc{LIGHT} program \citep{hill1979}. They concluded that both components are normal zero-age-main-sequence (ZAMS) stars of spectral types A1-A1.5 and F8-G5 with a chemical composition of $X~0.66, Y~0.3, Z~0.04$. 

However, summarizing the peculiarities of EK~Cep found in previous studies, \citet{popper1987} proposed that the secondary is a PMS star that is still contracting towards the ZAMS. In a follow-up spectroscopic study, \citet{martin1993} measured the abundances of Ca, Si and Fe and found them consistent with solar metallicity, with a surface lithium depletion value typical for very young stars. Based on these measurements, \citet{claret1995} uncovered additional evidence for the PMS nature of the secondary component through evolutionary modelling. They estimated the age of the system to be 20 Myr, with a primary that has just started hydrogen burning, and a secondary that is still contracting towards the main sequence. \citet{young2001} and \citet{marques2004} also found that the properties of the secondary in EK~Cep are well-matched with models of PMS stars; \citet{marques2004} refined the age estimate of the system to 26.8 Myr.

\citet{yildiz2003} constructed a stellar model with a core rotating up to 65 times faster than the envelope to explain the discrepancy between the properties of the primary and the secondary components, which could not be fitted with a single isochrone, but \citet{claret2006} showed that the discrepancy can be resolved without a rapidly rotating core both with standard and rotating stellar models.

\citet{landin2009} applied stellar models that account for tidal and rotational distortions to EK~Cep and reproduced the various measurements (radii, temperature ratio, rotation rates, lithium depletion and apsidal motion rate) for a much earlier age than the previous studies, between 15.5 and 16.7 Myr.

The most recent observational study of EK~Cep was done by \citet{anton2009}, who recorded photomultiplier light curves in UBVRI filters and performed polarimetric observations. They find significant variability in the polarization parameters of the system and ascribe it to the surface magnetic activity of the PMS secondary. The light curves in their study show no secondary minimum and appear to have a total eclipse in the primary minimum. However, the totality is poorly justified with few and highly scattered measurements. Our light curves show clear evidence of non-total eclipses, as can be seen in Fig. \ref{figEKCepLC}.

\subsection{HS~Her}
\label{secHSHerLit}

The first spectroscopic study of HS~Her was done by \citet{cesco45}, who were able to detect only the brighter component and classified it as a main sequence star of a type between B5 and B8.
After \citet{hall67} noted a large period variation based on the available eclipse timings, \citet{hall71} presented the first multicolour light curves and estimated the absolute parameters with a Fourier rectification method. They argued in favour of apsidal motion as the origin of the period change; according to the earliest estimates (disproved later), the period of the apsidal motion was only 15.5 years---the shortest known at the time. \citet{hall71} discuss at length a photometric anomaly in the trailing wing of the secondary eclipse and suggest the existence of opaque material inside the critical Roche lobe of the secondary---possibly the remnants from the protostellar phase, with the estimated age of the system at 20 Myr. This anomaly isn't present in our data (see Fig. \ref{figHSHerLC}).

\citet{giuricin81} revised the photometric elements of HS~Her using Wood's triaxial ellipsoids model \citep{wood1973}. Their results largely agreed with the ones from the previous study. \citet{wolf2002} confirmed the detection of apsidal motion, but with a period of 78 years, and proposed a third body in an eccentric orbit with a period of 85 years. The third body hypothesis was challenged by \citet{colak2005}, who cautioned that the only points on the O-C diagram that deviate from the apsidal motion fit are those made by visual observation and that they shouldn't be given the same consideration as the photographic, photoelectric and CCD measurements made later. Despite this argument, the disputed visual observations were repeatedly used in later O-C studies as indicative of a third body. We discuss this matter with regards to our own O-C analysis in Section \ref{secET}.

\citet{bozkurt2006} presented an updated analysis of HS~Her, based on UBV photoelectric observations and using the Wilson-Devinney method \citep[see e.g.,][]{wd2020ascl}. Their O-C analysis yields an apsidal motion period of 80.7 years and a third body in an eccentric orbit with a period of 85.4 years. They obtained the following system parameters: $T_1 = 15200 K$, $T_2 = 7600 K$, $i = 88.1^{\circ}$, $M_1 = 6.0 M_{\odot}$, $M_2 = 1.8 M_{\odot}$, $R_1 = 3.1 R_{\odot}$ and $R_2 = 1.7 R_{\odot}$.
\citet{kali2006} arrive to essentially the same light elements using their own photoelectric UBV measurements. Judging by the colour indices, they suggest that the secondary component of HS~Her appears to be of an earlier spectral type than it really is due to the irradiation from the primary, and that its intrinsic type is A7, in the PMS phase, rather than A4 on the main sequence as suggested by earlier studies. They estimate the age of the system at 17 Myr.

The most recent study of HS~Her was done by \citet{cakirli2007}, who measured, for the first time, the radial velocities of the faint secondary. This allowed them to construct a model combining their double-lined orbital solution with the results of previous photometric studies. Among other results, they derive the radial velocity curve of the hypothetical third body using the orbital elements from \citet{bozkurt2006}. However, they fail to detect a tertiary component in the spectra. The variability of the equivalent widths of Mg$_{II}\lambda$4481\angstrom\ and H$_{\alpha}$ lines is proposed as evidence of remaining protostellar material around the binary. Both components of HS~Her are estimated to be on the ZAMS, with near-solar metallicity and the age of 32~Myr.

Table \ref{tableGaia} lists the photometric magnitudes, color indices, parallaxes/distances and effective temperatures of EK~Cep and HS~Her, as provided in Gaia DR2 and EDR3. Note that the measurements are not necessarily taken during eclipses or in phases of maximum brightness, and therefore they cannot be used as reference for the effective temperature (through color index) or absolute magnitude (and therefore distance) calculations. They rather provide a "mean" temperature of the system, which does not describe the effective temperature of each component and does not match the spectroscopic observations, as we will show in the following paragraphs.

\begin{table}
    \centering
    \caption{Photometric magnitudes, color indices, parallaxes/distances and effective temperatures of EK~Cep and HS~Her, as provided in Gaia DR2 and EDR3.}
    \label{tableGaia}
    \begin{tabular}{lrr} 
        \hline
        Quantity    &   EK~Cep  & HS~Her    \\
        \hline        
        G [mag]     &   7.85    &   8.53    \\
        Bp [mag]    &   7.88    &   8.54    \\
        Rp [mag]    &   7.76    &   8.45    \\
        Bp-Rp [mag] &   0.12    &   0.06    \\
        $\pi$ [mas] &   5.81(2) &   2.03(2) \\
        d [pc]      &   172(1)  &   492(5)  \\
        T [K]       &   8635    &   9231    \\
        \hline
    \end{tabular}
\end{table}

\section{Photometric observations}
\label{secPhoto}

Photometric observations of EK~Cep and HS~Her were performed from the University of Athens Observatory (UOAO\footnote{\url{http://observatory.phys.uoa.gr}}). The facilities at UOAO utilize a robotic and remotely controlled Cassegrain reflector with diameter 0.4~m and focal ratio f/8 \citep{Gazeas2016}. An SBIG~ST10~XME CCD camera with a 2148$\times$1472 pixels chip was used in all observations. The telescope and camera, in combination with an f/6.3 focal reducer, resulted on a 17$\times$26 arcmin field of view. The camera is equipped with a set of BVRI (Bessell) filters, which provided multicolour photometric measurements in four optical bands.

EK~Cep was observed for a total of 52 nights between November 24, 2016 and March 26, 2017, providing 18313, 18186, 19074, 18660 photometric measurements in BVRI bands, respectively. 
The star GSC~4466:1166 (SAO~19575) (RA: 00$^{h}$ 13$^{m}$ 5$^{s}$.038, Dec: +05$^{\circ}$ 37$^{'}$ 53$^{''}$.08, G0V sp. type) was used for comparison, while the star GSC~4466:2583 (RA: 00$^{h}$ 13$^{m}$ 3$^{s}$.799, Dec: +05$^{\circ}$ 34$^{'}$ 55$^{''}$.63, K2V sp. type) was used as check star. 

HS~Her was observed for a total of 27 nights between May 20, 2017 and July 5, 2017, providing 5500, 4918, 5542 and 5915 photometric measurements in BVRI bands, respectively.
The star GSC~2113:1713 (RA: 00$^{h}$ 13$^{m}$ 5$^{s}$.038, Dec: +05$^{\circ}$ 37$^{'}$ 53$^{''}$.08, F2V sp. type) was used for comparison, while the star GSC~2113:2133 (SAO~86519) (RA: 00$^{h}$ 13$^{m}$ 3$^{s}$.799, Dec: +05$^{\circ}$ 34$^{'}$ 55$^{''}$.63, B9V sp. type) was used as check star. None of the comparison and check stars showed any variability through the entire observing season, while they have similar spectral type with the variables.

All digital images were reduced with dark and flat frames, which were acquired during each observing night. The instrumental and differential magnitudes of the stars were extracted using aperture photometry, while the aperture was adjusted according to the local seeing conditions. 

In addition to these observations, high-precision photometric data have been retrieved from TESS space mission \citep{tess}. HS~Her was observed only in long cadence mode (sector 26), while EK~Cep was observed in both short and long cadence mode (sectors 17, 18, 24 and 25).

The orbital PDCSAP data were retrieved via the MAST database\footnote{\textit{https://mast.stsci.edu/portal/Mashup/Clients/Mast/Portal.html}} and TESS-SPOC pipeline. The data were detrended by using the Savitzky-Golay filter \citep{savitzky1964} and normalized, phased and sigma-clipped for all prominent outliers.

\begin{figure*}
\includegraphics[width=\textwidth]{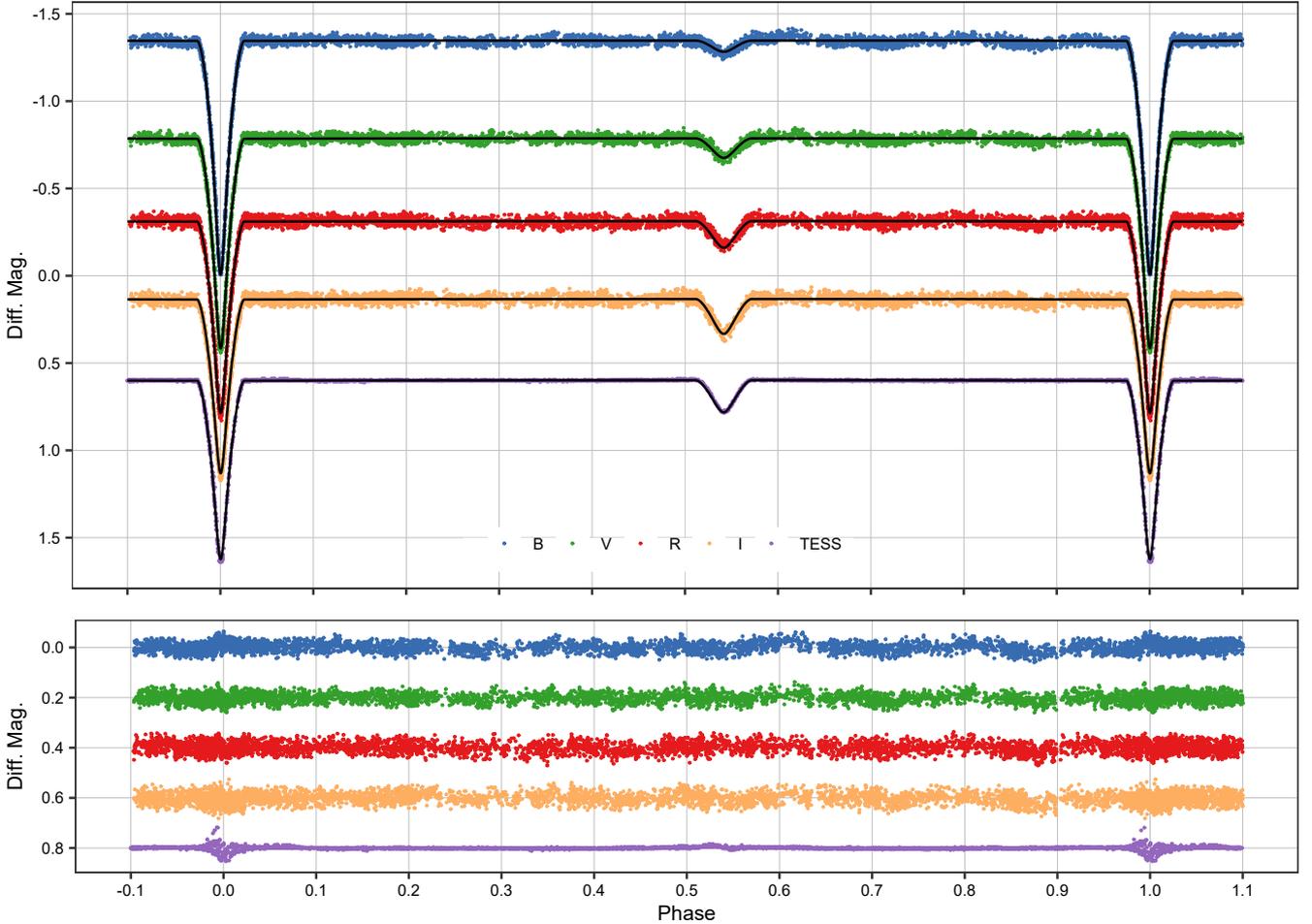}
\caption{The light curves of EK~Cep (a sample of 5000 points) with the model detailed in Table
\ref{tableModels} (solid line) and residuals (in the bottom panel).}
\label{figEKCepLC}
\end{figure*}

\begin{figure*}
\includegraphics[width=\textwidth]{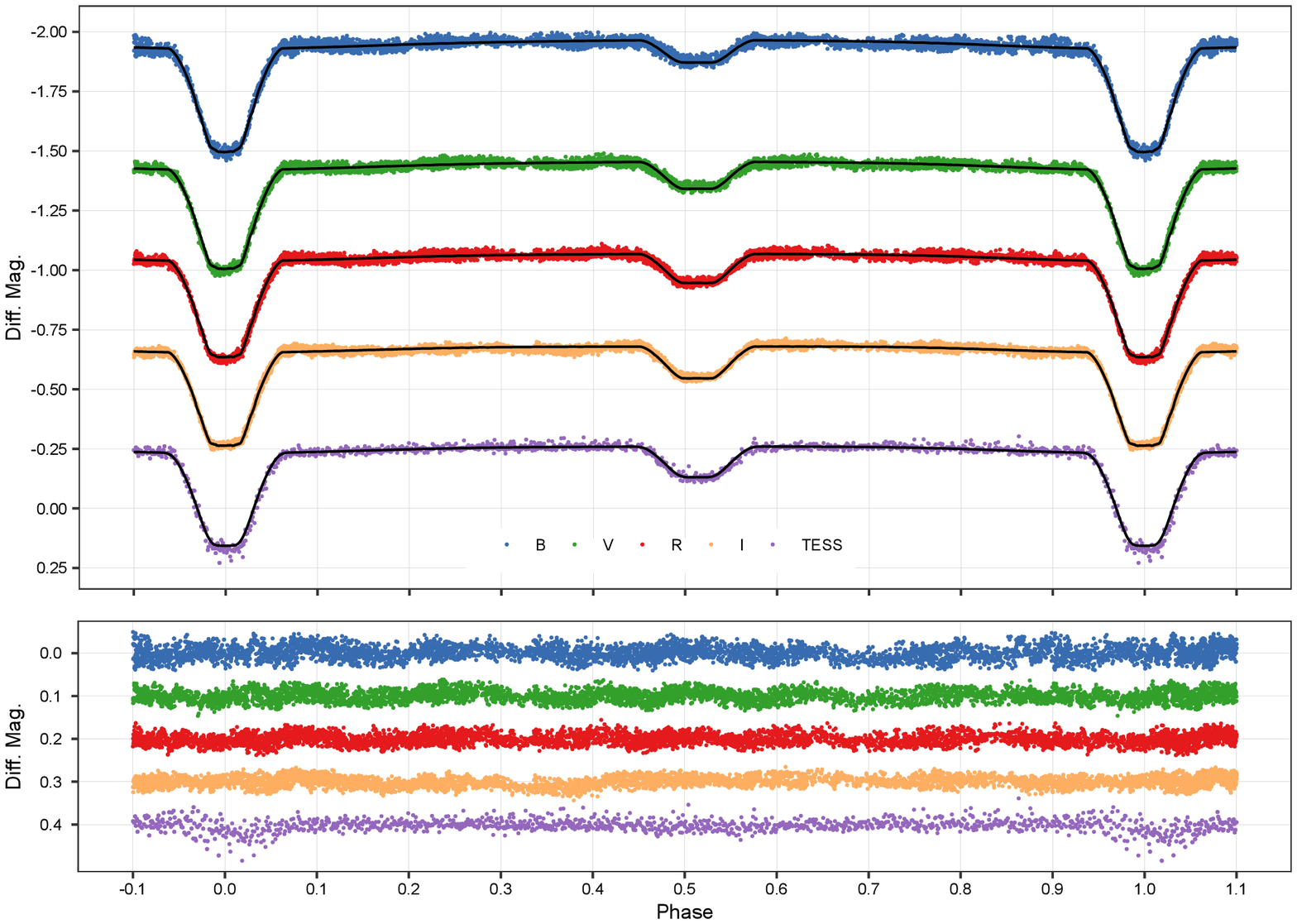}
\caption{The light curves of HS~Her (a sample of 5000 points) with the model detailed in Table \ref{tableModels} (solid line) and residuals (in the bottom panel).}
\label{figHSHerLC}
\end{figure*}

The light curves of EK~Cep and HS~Her are shown in Figs. \ref{figEKCepLC} and \ref{figHSHerLC}, and our ground-based observations are tabulated in Tables \ref{tableEkCepLC} and \ref{tableHsHerLC}, respectively.

\begin{table}
    \centering
    \caption{The light curves of EK~Cep. This table is available in its entirety in the machine-readable format. Only a portion is shown here for guidance regarding its form and content.}
    \label{tableEkCepLC}
    \begin{tabular}{lrrr} 
        \hline
        Filter & JD & Mag & Error \\
        \hline									
        B & 2457717.15538 & -1.334 & 0.009 \\
        B & 2457717.15601 & -1.350 & 0.009 \\
        B & 2457717.15664 & -1.351 & 0.008 \\
        ... \\
        I & 2457838.66156 & -0.404 & 0.022 \\
        I & 2457838.66208 &	-0.415 & 0.005 \\
        I & 2457838.66260 & -0.400 & 0.005 \\
        \hline									
    \end{tabular}
\end{table}

\begin{table}
    \centering
    \caption{The light curves of HS~Her. This table is available in its entirety in the machine-readable format. Only a portion is shown here for guidance regarding its form and content.}
    \label{tableHsHerLC}
    \begin{tabular}{lrrr} 
        \hline
        Filter & JD & Mag & Error \\
        \hline									
        B & 2457894.33244 &	-1.958 & 0.007 \\
        B & 2457894.33331 &	-1.965 & 0.007 \\
        B & 2457894.33418 &	-1.983 & 0.006 \\
        ... \\
        I & 2457939.59639 &	-0.979 & 0.007 \\
        I & 2457939.59703 &	-0.973 & 0.008 \\
        I & 2457939.59768 &	-0.949 & 0.015 \\
        \hline									
    \end{tabular}
\end{table}

\section{Spectroscopic observations}
\label{secSpectro}

High-resolution ($R \approx 15000$) spectroscopic observations were made using the 2-m Ritchey-Chretien telescope and the Coude horizontal spectrograph at NAO Rozhen, in two spectral regions, each about 200 Angstroms wide: in the vicinity of the Mg$_{II}$ line at 4481\angstrom\ (hereafter, the ``MgII4481'' region) and around the D1 and D2 sodium lines near 5900\angstrom\ (hereafter, the ``NaD5895'' region). 

For EK~Cep, 50 science frames were made during 6 observing sessions in 2005 and 2007; for HS~Her, 68 science frames were made over 11 observing sessions between 2002 and 2007. Each image was accompanied by the bias, dark and flat calibration frames. A thorium-argon lamp was used to record comparison spectra for wavelength calibration. 

\begin{figure*}
\includegraphics[width=\textwidth]{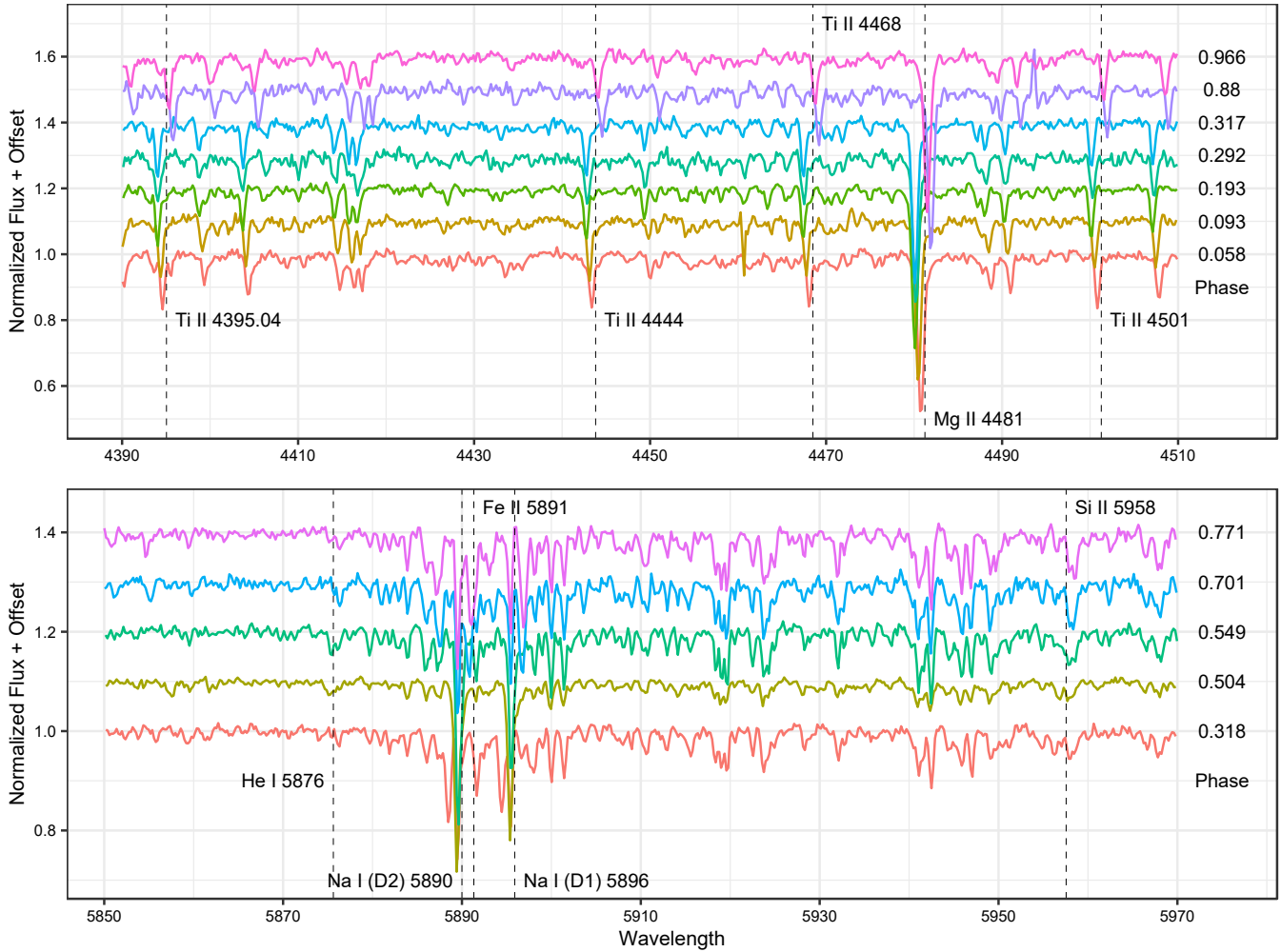}
\caption{A selection of spectra for EK~Cep.}
\label{figEKCepSpectra}
\end{figure*}

\begin{figure*}
\includegraphics[width=\textwidth]{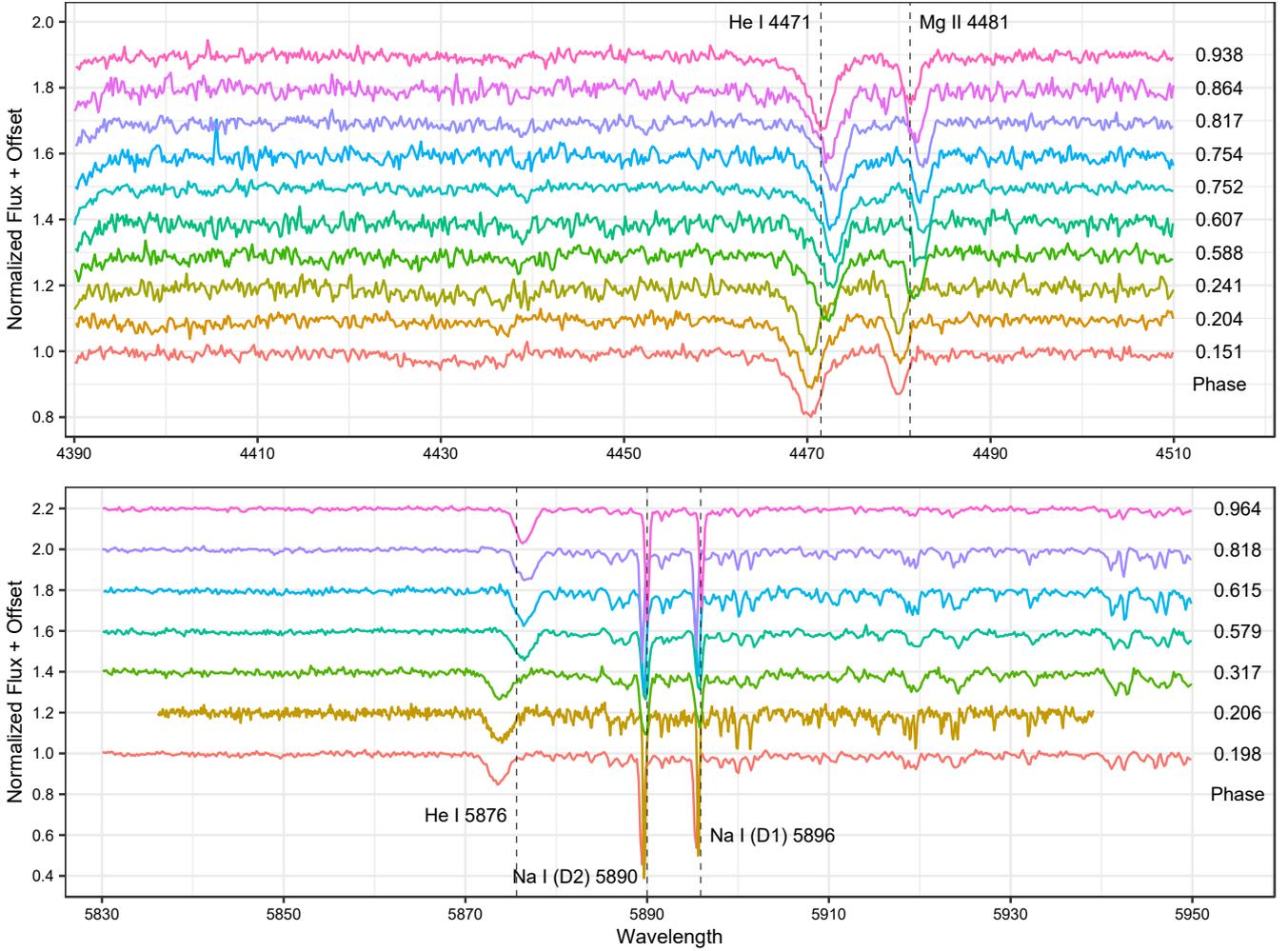}
\caption{A selection of spectra for HS~Her.}
\label{figHSHerSpectra}
\end{figure*}

We reduced all the science frames in \textsc{iraf}\footnote{\textit{http://ast.noao.edu/data/software}} \citep{iraf}, following the standard CCD processing steps. Examples of the reduced spectra at different orbital phases (about one-third of the entire sample) with provisional line identification are shown in Figs. \ref{figEKCepSpectra} and \ref{figHSHerSpectra}. 

\begin{figure}
\includegraphics[width=0.495\textwidth]{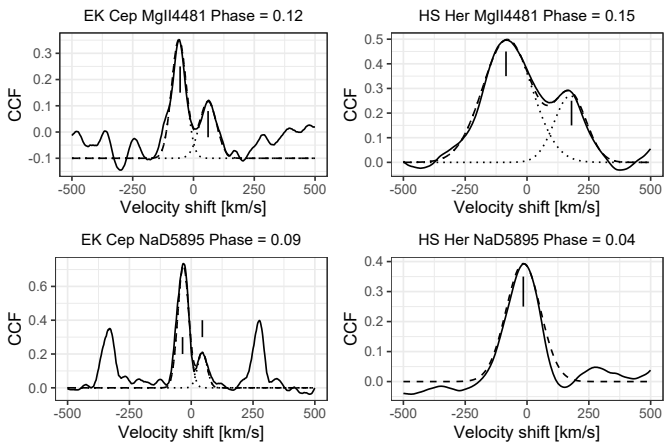}
\caption{A sample of cross-correlation functions for EK~Cep (on the left) and HS~Her (on the right) in the  MgII4481 (top) and NaD5895 region (bottom). The higher peak corresponds to the primary component in all plots. The lower-right plot demonstrates the situation where the secondary component could not be detected.}
\label{figCCF}
\end{figure}

The radial velocities were measured with the cross-correlation function (CCF) method of \citet{tonry1979} as implemented in the \textsc{fxcor} task included with the \textsc{IRAF} RV package\footnote{\textit{http://iraf.noao.edu/projects/rv/rv.html}}. As templates, we used a grid of synthetic spectra calculated with the spectral synthesis program \textsc{spectrum}\footnote{\textit{http://www.appstate.edu/\%7Egrayro/spectrum/spectrum.html}} \citep{gray1994} covering  effective temperatures in the range from 5000 to 15000 K, surface gravity logarithms in the range from 3.5 to 4.5 and rotational broadening velocities between 0 and 200 km/s (for a total of about 4500 templates). Templates close to the expected parameters of the target stars' components were then hand-picked according to the quality of the resulting CCF. In the case of EK~Cep, the best results could be achieved with templates at T = 9000 K, $\log g$ = 4.5 and $v \sin i$ between 20 and 40 km/s for the primary, and T = 5750 K, $\log g$ = 4.5 and $v \sin i$ between 0 and 20 km/s for the secondary component. The rotational velocities of the components at the equator, assuming synchronous rotation and based on the radii reported in Table \ref{tableAbsPars}, are $v_1 \approx 18$ km/s and $v_2 \approx 15 km/s$, in broad agreement with the best-fitting templates. For HS~Her, the templates at T = 15000 K, $\log g$ = 4.5 and $v \sin i$ between 70 and 90 km/s for the primary, and T = 7750 K, $\log g$ = 4.5 and $v \sin i$ between 20 and 40 km/s for the secondary component were found to provide the best measurements. The estimated rotational velocities are, again, in broad agreement with this selection, at $v_1 \approx 81$ km/s and $v_2 \approx 46 km/s$. A sample of CCFs is given in Fig. \ref{figCCF}.

No evidence of a possible tertiary component could be seen in the spectra or the CCFs of either star.

\begin{figure*}
\includegraphics[width=0.495\textwidth]{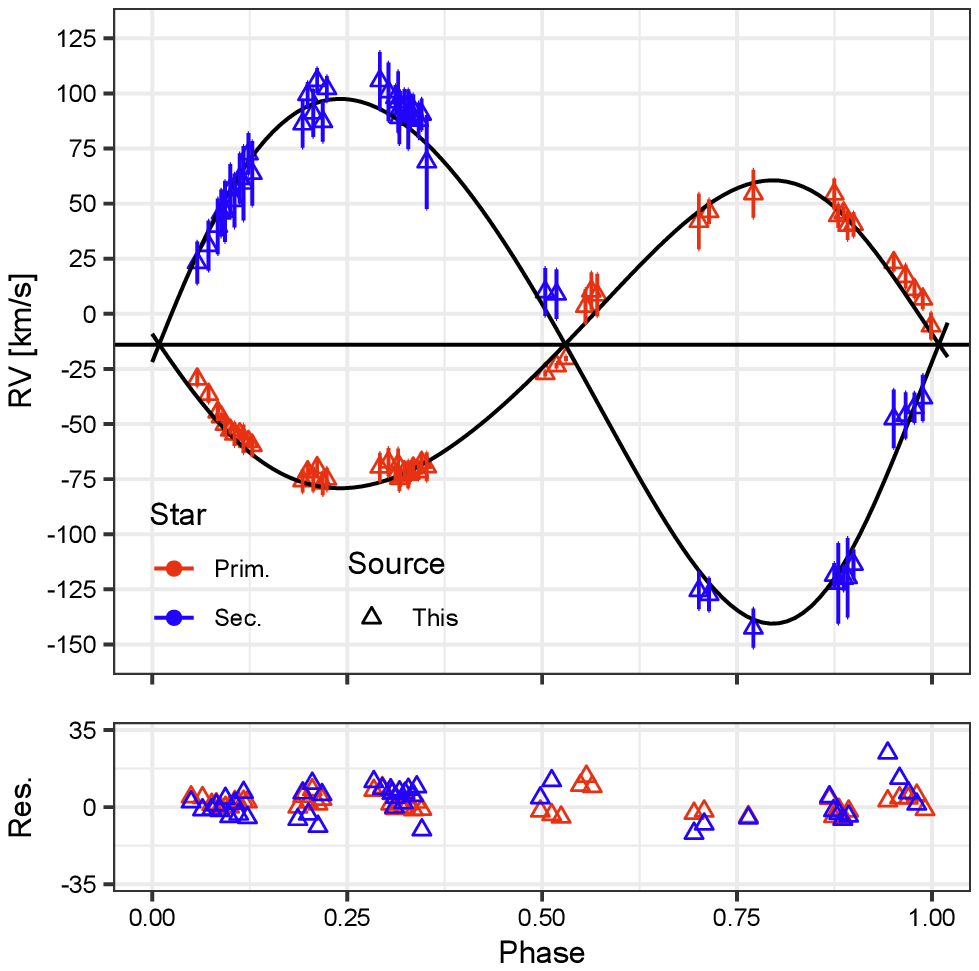}
\includegraphics[width=0.495\textwidth]{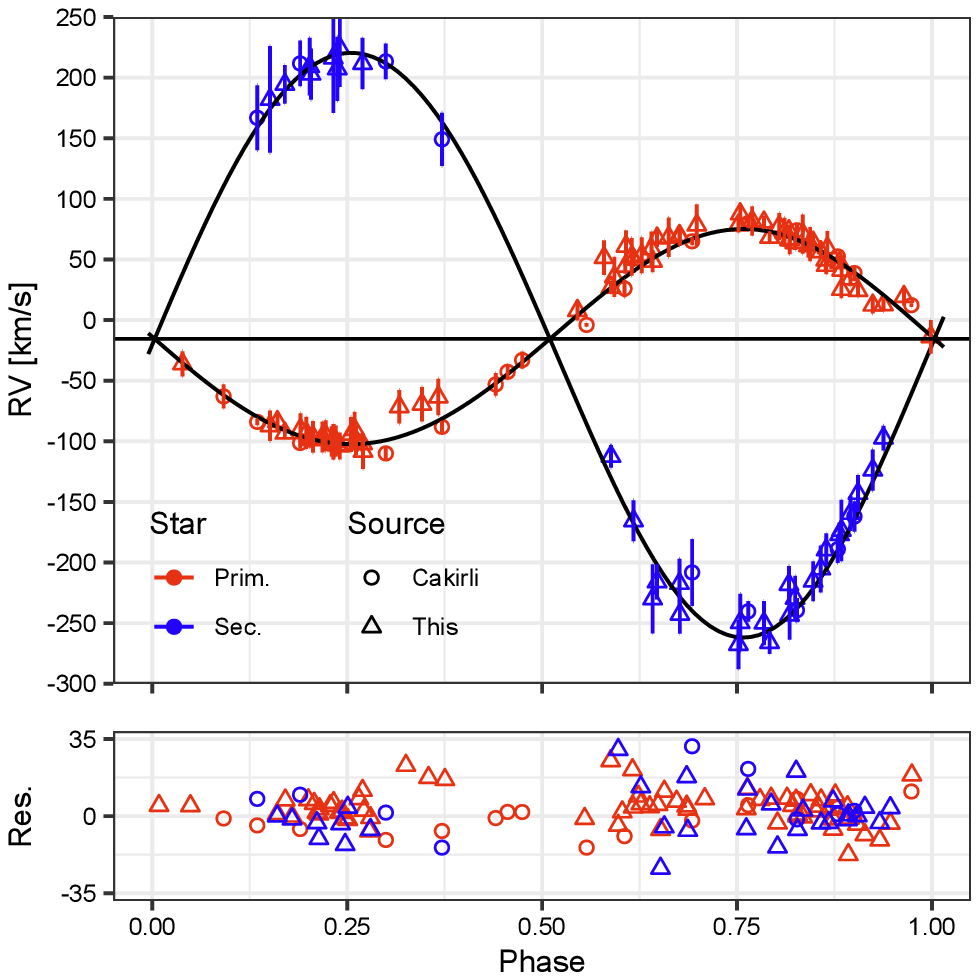}
\caption{The RV curves of EK~Cep (left panel) and HS~Her (right panel) with the preliminary models and residuals.}
\label{figRV}
\end{figure*}

The radial velocities are tabulated in Tables \ref{tabEKCepRVData} and \ref{tabHSHerRVData}, and shown in Fig. \ref{figRV}. The average error of RV measurements for the primary (secondary) component of EK~Cep is 5 (10) km/s; and for HS~Her, 10 (20) km/s.

In addition to these new measurements, we use the radial velocities measured by \citet{cakirli2007} for the orbital solution of HS~Her. The other previous radial velocity measurements (i.e. those from \citealt{cesco45} for HS~Her, \citealt{tomkin1983} and \citealt{ebbi1966b} for EK~Cep) were published without measurement errors and could not be analyzed in a consistent way together with our data.

\begin{table*}    
\caption{Radial velocity measurements for EK~Cep. This table is available in its entirety in the machine-readable format. Only a portion is  shown here for guidance regarding its form and content.}    
\label{tabEKCepRVData}
\begin{center}
\begin{small}
\begin{tabular}{llrrrrrr}
\hline
\hline
Sp. region & BJD & RV$_1$ & $\Delta$RV$_1$ & RV$_2$ & $\Delta$RV$_2$ & Exp. time & SNR \\ 
& (mid-exposure) & [km/s] & [km/s]         & [km/s] & [km/s]         & [s]       &     \\
\hline
MgII4481 & 2453539.398006 & -29.57 & 3.39 & 23.23 &  9.50 & 1200 & 114.2 \\
MgII4481 & 2453539.462715 & -36.53 & 3.71 & 30.94 & 11.33 & 1200 & 138.3 \\
MgII4481 & 2453539.515131 & -44.57 & 3.97 & 39.70 & 12.60 & 1200 & 132.0 \\
... \\
NaD5895  & 2454253.432761 & -74.71 & 3.86 & 93.28 & 7.74  & 1200 & 155.4 \\
NaD5895  & 2454253.487938 & -72.38 & 3.89 & 92.02 & 7.93  & 1200 & 173.2 \\
NaD5895  & 2454253.536367 & -71.67 & 3.49 & 87.76 & 8.09  & 1200 & 185.3 \\
\hline
\end{tabular}
\end{small}
\end{center}
\end{table*}

\begin{table*}    
\caption{Radial velocity measurements for HS~Her. This table is available in its entirety in the machine-readable format. Only a portion is  shown here for guidance regarding its form and content.} 
\label{tabHSHerRVData}
\begin{center}
\begin{small}
\begin{tabular}{lrrrrrrr}
\hline
\hline
Sp. region & BJD & RV$_1$ & $\Delta$RV$_1$ & RV$_2$ & $\Delta$RV$_2$ & Exp. time & SNR \\ 
& (mid-exposure) & [km/s] & [km/s]         & [km/s] & [km/s]         & [s]       &     \\
\hline
MgII4481 & 2453157.478163 &   33.78 & 5.75 & -159.53 & 14.39 & 1200 & 84.1 \\
MgII4481 & 2453157.547568 &   12.65 & 5.32 &  -97.29 & 10.11 & 1200 & 77.7 \\
MgII4481 & 2453180.351405 &   45.06 & 7.18 & -189.59 & 13.64 & 1200 & 52.1 \\
... \\
NaD5895  & 2452478.450399 & - 96.41 & 12.99 &        &       & 1200 & 74.6 \\
NaD5895  & 2452478.498028 &  -99.93 & 12.75 &        &       & 1200 & 60.0 \\
NaD5895  & 2452478.555418 & -108.24 & 14.51 &        &       & 1200 &      \\
\hline
\end{tabular}
\end{small}
\end{center}
\end{table*}

\section{Eclipse timing variations}
\label{secET}

EK~Cep and HS~Her both have a history of observations and eclipse timings several decades long. 
We gathered all the eclipse timings available in the literature and in online databases. In addition, we extracted nine more new times of minimum light for EK~Cep and four more  for HS~Her, using the method of \citet{Kwee1956}, from our new ground-based observations. They are the average values from BVRI filters, and the reported uncertainties are also average values of individual measurements. Another 38 eclipse timings were extracted from the TESS data for EK~Cep and ten more for HS~Her. The eclipse times newly measured in this work are given in Table \ref{tableTOM}.

\begin{figure*}
\includegraphics[width=0.495\textwidth]{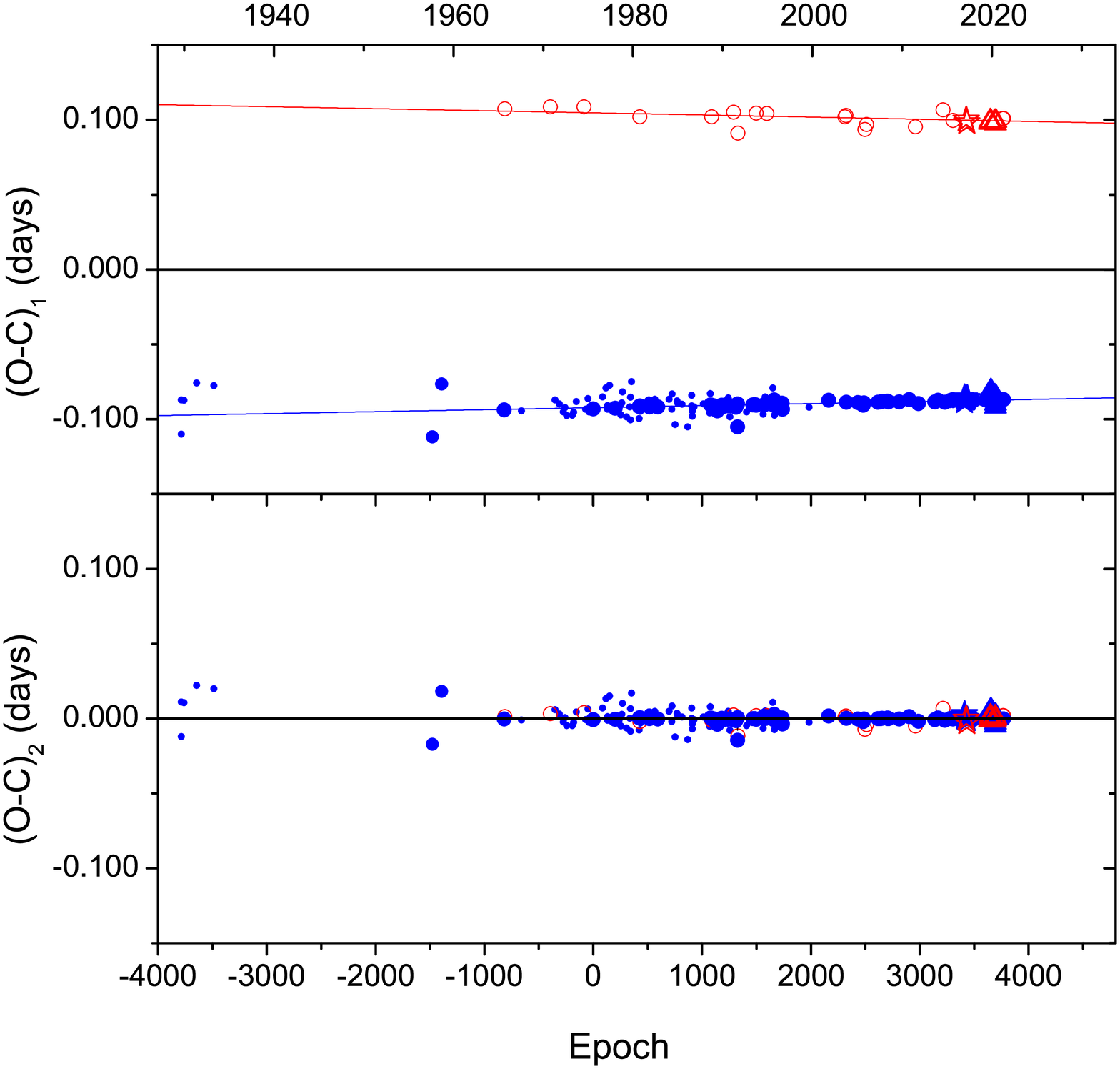}
\includegraphics[width=0.495\textwidth]{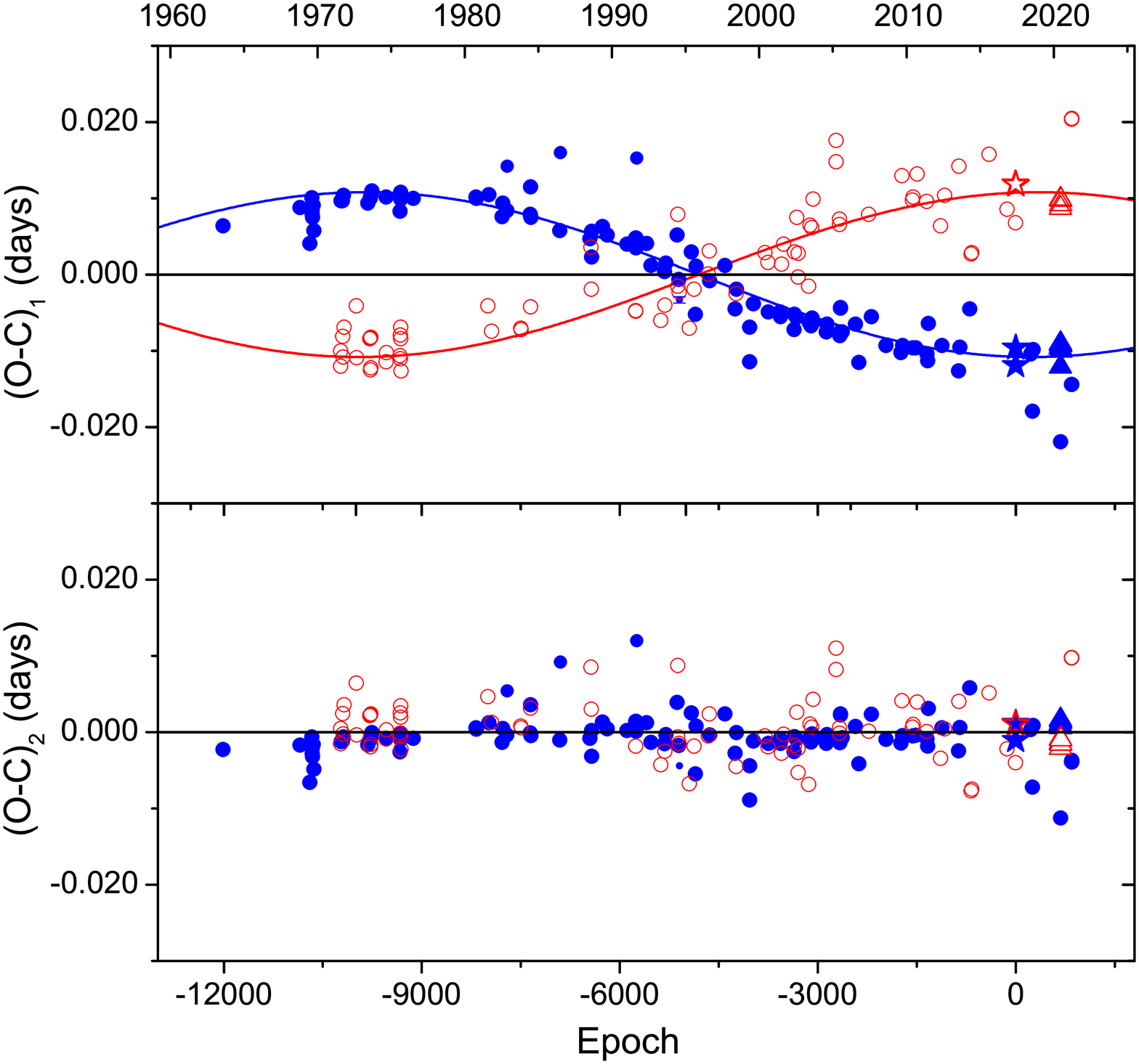}
\caption{The O-C diagrams for EK~Cep (left) and HS~Her (right). The primary and secondary eclipse timings are denoted with filled and open symbols, respectively. Eclipse timings collected from literature are marked with circles, whose size is proportional to the assigned weight. Measurements obtained from our ground-based observations are marked with stars, while those from TESS observations are marked with triangles.}
\label{figOC}
\end{figure*}

All the data are used to construct the $O-C$ diagrams, which are excellent tools for orbital investigations and provide valuable information about secular changes of the orbital period. The light-time effect (LITE) may also provide an indication for the existence of a third body orbiting the binary. The $O-C$ diagrams for EK~Cep and HS~Her are shown in Fig. \ref{figOC}. Both stars clearly present an eccentric orbit with apsidal motion, reflected by the separation of the primary and secondary minima. 

The linear ephemerides resulting from the $O-C$ analysis are given in Eqs. \ref{EqEKCepEph} and \ref{EqHSHerEph}.

\begin{equation}
    \rm EK\ Cep: 
    Min\ I = 2457732.29816(23) + 4.427794(5) \times E
    \label{EqEKCepEph}
\end{equation}

\begin{equation}
    \rm HS\ Her: 
    Min\ I = 2457909.47190(18) + 1.6374352(2) \times E
    \label{EqHSHerEph}
\end{equation}

We model the eclipse timing variations using the \textsc{LITE} software \citep{zasche2009}. Starting from the basic $O-C$ expression (Eq. \ref{EqOC1}), we include the apsidal motion term (Eq. \ref{EqOC2}) from the mathematical formulation given in \citet{martynov1973}. 


\begin{equation}
    \centering
      (O-C)_1 = T - (T_{0} + P \times E )
    \label{EqOC1}
\end{equation}

\begin{equation}
    \centering
      (O-C)_2 = (O-C)_1 \pm \frac{Pe}{2\pi}(1+\frac{1}{sin^{2}i})cos(\omega_{0}+\dot{\omega}E)
    \label{EqOC2}
\end{equation}


Statistical weights were assigned to the input eclipse timings according to the method of observation and reliability. We used the values 1, 5, and 10 for visual, photographic and CCD observations, respectively. In the case of HS~Her, the highly scattered visual eclipse timings taken before 1960 were discarded. 

The obtained orbital parameters are listed in Tables \ref{tableocekcep} and \ref{tableochsher}, together with the results of $O-C$ analyses performed in previous studies. 
Our analysis gives no indication of a third member orbiting either of the two systems. Regarding the tertiary component hypothesized in previous studies of HS~Her, the eclipse timings gathered during the last two decades neatly confirm the predictions of a pure apsidal motion model. Compare, for example, the $O-C$ diagram for HS~Her in Fig. \ref{figOC} with Fig. 1 from \citet{wolf2002}. Now that almost the entire apsidal motion period has been covered with observations, little uncertainty is left regarding the nature and parameters of the eclipse timing variations. Based on these measurements and in line with the spectroscopic evidence, the hypothesis of the third body in HS~Her can be rejected.

Note that the uncertainty of orbital parameters in older studies seems to be underestimated, or it is not given at all. Significant underestimation is prominent in the case of EK~Cep, where the $O-C$ diagram does not provide any evidence of fast apsidal motion. Therefore, although some of the orbital parameters in our study are provided with large uncertainties, they describe the observed orbital motion in a more realistic and robust way.

\begin{table}
\centering
\caption{New times of minimum light (primary and secondary eclipses) for EK~Cep and HS~Her
measured from our ground-based observations and the TESS data.}
\label{tableTOM}
\begin{scriptsize}
\begin{tabular}{lllcc} 
\hline
JD (days)	&	error	&	type	&	filter	&	Observatory	\\
\hline	
EK~Cep	        &		    &		&		    &		    \\
\hline									
2457732.29959	&	0.00023	&	I	&	BVRI	&	UOAO	\\
2457741.15759	&	0.00035	&	I	&	BVRI	&	UOAO	\\
2457785.43292	&	0.00010	&	I	&	BVRI	&	UOAO	\\
2457794.28826	&	0.00013	&	I	&	BVRI	&	UOAO	\\
2457805.54403	&	0.00046	&	II	&	BVRI	&	UOAO	\\
2457807.57135	&	0.00009	&	I	&	BVRI	&	UOAO	\\
2457816.42675	&	0.00011	&	I	&	BVRI	&	UOAO	\\
2457836.53609	&	0.00024	&	II	&	BVRI	&	UOAO	\\
2457838.56562	&	0.00008	&	I	&	BVRI	&	UOAO	\\
2458766.37408	&	0.00157	&	II	&	C	    &	TESS	\\
2458768.40480	&	0.00163	&	I	&	C	    &	TESS	\\
2458770.80260	&	0.00111	&	II	&	C	    &	TESS	\\
2458772.83213	&	0.00044	&	I	&	C	    &	TESS	\\
2458775.23084	&	0.00070	&	II	&	C	    &	TESS	\\
2458779.65862	&	0.00233	&	II	&	C	    &	TESS	\\
2458781.68788	&	0.00232	&	I	&	C	    &	TESS	\\
2458784.08687	&	0.00181	&	II	&	C	    &	TESS	\\
2458786.11559	&	0.00088	&	I	&	C	    &	TESS	\\
2458792.94220	&	0.00115	&	II	&	C	    &	TESS	\\
2458794.97180	&	0.00146	&	I	&	C	    &	TESS	\\
2458797.37012	&	0.00147	&	II	&	C	    &	TESS	\\
2458799.39836	&	0.00187	&	I	&	C	    &	TESS	\\
2458803.82607	&	0.00227	&	I	&	C	    &	TESS	\\
2458806.22498	&	0.00284	&	II	&	C	    &	TESS	\\
2458808.25956	&	0.00207	&	I	&	C	    &	TESS	\\
2458810.65308	&	0.00257	&	II	&	C	    &	TESS	\\
2458812.68179	&	0.00136	&	I	&	C	    &	TESS	\\
2458956.77032	&	0.00151	&	II	&	C	    &	TESS	\\
2458958.80005	&	0.00315	&	I	&	C	    &	TESS	\\
2458961.19772	&	0.00159	&	II	&	C	    &	TESS	\\
2458963.22737	&	0.00138	&	I	&	C	    &	TESS	\\
2458965.62578	&	0.00120	&	II	&	C	    &	TESS	\\
2458967.65440	&	0.00337	&	I	&	C	    &	TESS	\\
2458970.05342	&	0.00192	&	II	&	C	    &	TESS	\\
2458972.08431	&	0.00759	&	I	&	C	    &	TESS	\\
2458974.48147	&	0.00031	&	II	&	C	    &	TESS	\\
2458976.51017	&	0.00242	&	I	&	C	    &	TESS	\\
2458978.90941	&	0.00196	&	II	&	C	    &	TESS	\\
2458980.93809	&	0.00045	&	I	&	C	    &	TESS	\\
2458985.36636	&	0.00900	&	I	&	C	    &	TESS	\\
2458989.79017	&	0.00499	&	I	&	C	    &	TESS	\\
2458992.19126	&	0.00621	&	II	&	C	    &	TESS	\\
2458994.21602	&	0.00550	&	I	&	C	    &	TESS	\\
2458998.64929	&	0.00404	&	I	&	C	    &	TESS	\\
2459001.04844	&	0.00081	&	II	&	C	    &	TESS	\\
2459005.47636	&	0.00169	&	II	&	C	    &	TESS	\\
2459007.50622	&	0.00137	&	I	&	C	    &	TESS	\\
\hline									
HS~Her	        &		    &	    &		    &		    \\
\hline									
2457905.40531	&	0.00034	&	II	&	BVRI	&	UOAO	\\
2457909.47519	&	0.00018	&	I	&	BVRI	&	UOAO	\\
2457914.38974	&	0.00095	&	I	&	BVRI	&	UOAO	\\
2457918.50471	&	0.00033	&	II	&	BVRI	&	UOAO	\\
2459011.47056	&	0.00373	&	I	&	C	    &	TESS	\\
2459018.85857	&	0.00133	&	II	&	C	    &	TESS	\\
2459019.65530	&	0.00277	&	I	&	C	    &	TESS	\\
2459021.29575	&	0.01082	&	I	&	C	    &	TESS	\\
2459023.76964	&	0.00049	&	II	&	C	    &	TESS	\\
2459024.57067	&	0.00160	&	I	&	C	    &	TESS	\\
2459031.95731	&	0.00257	&	II	&	C	    &	TESS	\\
2459032.75682	&	0.00187	&	I	&	C	    &	TESS	\\
2459033.59535	&	0.00177	&	II	&	C	    &	TESS	\\
2459034.39497	&	0.00051	&	I	&	C	    &	TESS	\\
\hline									
\end{tabular}
\end{scriptsize}
\end{table}

\begin{table*}
\centering
\caption{The orbital parameters of EK~Cep from past research and in this work.}
\label{tableocekcep}
\begin{tabular}{rcccccc} 
\hline
\hline 
Ref & $T_{0} [JD]$ & $P_{orb}\ [d]$ & e & $\omega\ [^{\circ}]$ & $d\omega/dt\ (deg/yr)$ & $U (yr)$ \\
\hline
$[1]$  & 2439002.722	   & 4.42775	    & 0.09	    & 33	    & -	           & -	        \\
$[2]$  & 2439002.282(24)   & 4.42775	    & 0.126(15)	& 49.8(7.2)	& -	           & -	        \\
$[3]$  & 2439072.51	       & 4.42775	    & 0.129	    & 47	    & -	           & -	        \\
$[4]$  & -	               & 4.427796	    & -	        & -	        & -	           & -	        \\
$[5]$  & -	               & 4.4277926(9)   & 0.09	    & 50	    & -	           & -	        \\
$[6]$  & -	               & 4.4278	        & -	        & -	        & -	           & -	        \\
$[7]$  & -	               & -	            & -	        & -	        & 0.076	       & 4700	    \\
$[8]$  & 2445161.354(19)   & 4.427822(29)   & 0.109(3)	& 46.9(1.4)	& -	           & -	        \\
$[9]$  & 2443519.074	   & 4.4277964      & 0.109	    & 49.84(14)	& 0.082(8)	   & 4400(400)	\\
$[10]$ & 2442624.75198(51) & 4.44278070(35) & 0.109	    & 50(2)	    & 0.088(26)	   & 4100(1217)	\\
$[11]$ & 2443519.077	   & 4.4277964      & -	        & -	        & -	           & -	        \\
$[12]$ & -	               & 4.4277954      & -	        & -	        & 0.086(4)     & 4200(200) 	\\
$[13]$ & 2442624.6590(3)   & 4.4277960(3)   & 0.109	    & 50(1.4)	& 0.083(12)    & 4300(638)	\\
$[14]$ & -	               & -	            & -	        & -	        & 0.081(7)     & 4460(16)	\\
$[15]$ & 2442624.752(9)	   & 4.427794(4)    & 0.109(14) & 50(3)     & 0.061(0.203) & 5860(6018)	\\
\hline
\multicolumn{7}{c}{$[1]$ \cite{ebbi1966a}; $[2]$ \cite{ebbi1966b}; $[3]$ \cite{lucy1971}; $[4]$ \cite{guarn1975};} \\
\multicolumn{7}{c}{$[5]$ \cite{koch1977}; $[6]$ \cite{gimenez1980}; $[7]$ \cite{khaliu1983}; $[8]$ \cite{tomkin1983};} \\
\multicolumn{7}{c}{$[9]$ \cite{hill1984}; $[10]$ \cite{gimenez1985}; $[11]$ \cite{ebbi1990}; $[12]$ \cite{zakirov1993};} \\
\multicolumn{7}{c}{$[13]$ \cite{claret1995}; $[14]$ \cite{yildiz2003}; $[15]$ this paper.}
\end{tabular}
\end{table*}

\begin{table*}
\centering
\caption{The orbital parameters of HS~Her from past research and in this work.}
\label{tableochsher}
\begin{tabular}{rccccccccc} 
\hline
\hline	
& $T_0\ [JD]$ & $P_{orb}\ [d]$ & e & $\omega\ [^{\circ}]$ & $d\omega/dt [^{|circ}/yr]$ & $U\ [yr]$ & $P_3\ [yr]$ & $e_3$ \\
\hline
$[1]$  &	-	            & 1.637416	    & 0.05(0.02) & 37(19)   & -	      &	-	    &	-	  & -	    \\
$[2]$  & 2430098.902	    & 1.637416	    & 0.069	     & 29	    & -	      &	-	    &	-	  &	-	    \\
$[3]$  & 2437854.194	    & 1.6374333	    & 0.033	     & 240	    & 23.2258 &	15.5    &	-	  &	-	    \\
$[4]$  &	-	            & -	            & 0.05	     & 37	    & -	      &	120	    &	-	  &	-	    \\
$[5]$  &	-	            & 1.63744	    & -	         &	-	    & -	      &	-	    &	-	  &	-	    \\
$[6]$  & 2440146.6008	    & 1.6374333	    & -	         &	-	    & -	      &	60	    &	-	  &	-	    \\
$[7]$  &	-	            & 1.637	        & -	         &	-	    & -	      &	92(14)	&	-	  &	-	    \\
$[8]$  &	-	            & 1.637	        & 0.0190(6)	 & -	    & -	      & 92(14)	&	-	  &	-	    \\
$[9]$  & 2447382.4104(3)	& 1.63743125(7)	& 0.020(3)	 & 232(3)   & 4.6(1)  & 78.0(3)	& 86(2)	  & 0.80(7) \\
$[10]$ & 2452417.5275       & 1.637438      & -	         &	-	    & -	      &	-	    &	-	  &	-	    \\
$[11]$ & 2452856.3646(2)	& 1.6374341(1)	& 0.0205(10) & 303(4)   & 4.7(2)  & 77(3)	&	-	  &	-	    \\
$[12]$ & 2447382.4224(2)	& 1.63743402(9)	& 0.0188(5)	 & 236(2)   & 4.3(2)  & 84(4)	&	?	  &	?	    \\
$[13]$ & 2453584.953	    & 1.6374316	    & 0.05(1)	 & 72(2)    & -       &	-	    &	-	  &	-	    \\
$[14]$ & 2447382.406(4)	    & 1.637432(5)	& 0.021(1)	 & 235(2)   & 4.5(2)  & 81(4)	& 85(9)	  & 0.90(8) \\
$[15]$ & 2447382.4062(45)	& 1.6374316(47)	& 0.020(1)	 & 236(2)   & 4.4(2)  & 82(4)	& 89(10)  & 0.91(8) \\
$[16]$ &	-	            & -	            & 0.048(1)	 & -	    & -	      & -	    &	-	  &	-	    \\
$[17]$ &	-	            & -	            & 0.059(1)	 & -	    & -	      & -	    &	-	  &	-	    \\
$[18]$ & 2452501.0424(9)    & 1.6374329(2)	& 0.021(2)	 & 305(1)   & 4.4	  &	70(5)	& 112(28) & -	    \\
$[19]$ & 2457909.487(1)	    & 1.6374342(2)	& 0.021(2)	 & 355.7(8)	& 4.0(3)  & 91(6)	& -	      & -       \\
\hline
\multicolumn{10}{c}{$[1]$ \cite{cesco45}; $[2]$ \cite{lucy1971}; $[3]$ \cite{hall71}; $[4]$ \cite{martynov1972};} \\
\multicolumn{10}{c}{$[5]$ \cite{bran1980}; $[6]$ \cite{todoran1992}; $[7]$ \cite{khaliu1992}; $[8]$ \cite{petrova1999};} \\
\multicolumn{10}{c}{$[9]$ \cite{wolf2002}; $[10]$ \cite{borkovits2002}; $[11]$ \cite{colak2005}; $[12]$ \cite{khaliu2006};} \\
\multicolumn{10}{c}{$[13]$ \cite{cakirli2007}; $[14]$ \cite{bozkurt2006}; $[15]$ \cite{bozkurt2007}; $[16]$ \cite{karami2008};} \\
\multicolumn{10}{c}{$[17]$ \cite{karami2009}; $[18]$ \cite{bulut2017}; $[19]$ this paper.}
\end{tabular}
\end{table*}

\section{Light and radial velocity curve modelling}
\label{secModeling}

The modelling of light and velocity curves was done by simultaneous fitting with \textsc{phoebe 1.0} \citep{phoebe}. For both binaries, we used the detached configuration. The albedos and gravity darkening coefficients were fixed at their theoretical values of $A = 0.5$ and $g = 0.32$ for stars with convective envelopes ($T < 7200$~K) and $A = 1$ and $g = 1$ for stars with radiative envelopes ($T > 7200$~K) \citep{rucinski1969,lucy1967,vonzeipel1924}. The limb darkening coefficients were taken from the tables of \citet{vanhamme1993AJ} according to the effective temperatures of the components and the filters used. For EK~Cep, we adopted the logarithmic, and for HS~Her, the square root limb darkening law. The asynchronous rotation parameter was assumed to be one. We tried, and rejected, larger values (corresponding to rotation faster than the orbital revolution) for both stars. Preliminary modelling similarly showed that adding spots and the third light to the models is not necessary for either star.

In the case of EK~Cep, the primary temperature was fixed to the value determined by \citet{popper1987}, $T_1=9000 K$; for HS~Her, to the value determined by \citet{cakirli2007}, $T_1=15200 K$. The mass ratio ($q$), inclination ($i$), orbital separation ($a$), eccentricity ($e$), argument of periastron ($\omega$), secondary temperature ($T_2$) and the surface potentials of both components ($\Omega_{1,2}$) were treated as free parameters.

The models are shown together with the observations in Figs. \ref{figEKCepLC}, \ref{figHSHerLC} and \ref{figRV}, and the results are summarized in Table \ref{tableModels}. The errors given there are the formal fitting errors reported by the software and available only for the adjustable parameters of the model. We discuss a more robust error estimation scheme in relation to the absolute parameters in the next section.

\begin{table*}    
\caption{The parameters and resulting quantities of simultaneous LC+RV modelling.}    
\label{tableModels}
\centering
\begin{tabular}{lrrrrrrrr}
\hline
                       & & & & EK~Cep & & & & HS~Her \\
\hline
\hline
Quantity               & & Value & Formal Err. & Estim. Err. & & Value     & Formal Err. & Estim. Err. \\
\hline
$T_1   [K]$            & & 9000     & -        & 200         & & 15200     & -           & 1000        \\
$T_2 [K]$              & & 5655     & 4        & 80          & & 7929      & 5           & 210         \\
$a [R_{\odot}]$        & & 16.44    & 0.7      & 0.02        & & 10.68     & 0.3         & 0.009       \\
$q = M_2/M_1$          & & 0.58668  & 0.0009   & 0.004       & & 0.36790   & 0.0009      & 0.001       \\
$\gamma [km/s]$        & & -14      & 2        & 0.08        & & -15.5     & 2           & 0.04        \\
$i [^{\circ}]$         & & 88.481   & 0.02     & 0.03        & & 87.764    & 0.02        & 0.02        \\
$\omega [^{\circ}]$    & & 53.03    & 0.2      & 0.1         & & 13.3      & 2           & 2           \\
$e$                    & & 0.10788  & 0.0004   & 0.0003      & & 0.02213   & 0.0003      & 0.0002      \\
$\Omega_1$             & & 11.438   & 0.01     & 0.006       & & 4.4544    & 0.004       & 0.02        \\
$\Omega_2$             & & 8.535    & 0.02     & 0.05        & & 3.9844    & 0.006       & 0.006       \\
\hline
$K_1 [km/s]$           & & 69.8     &          & 0.3         & & 88.7      &             & 0.3         \\
$K_2 [km/s]$           & & 119.1    &          & 0.9         & & 241.2     &             & 0.8         \\
\hline
\hline
                       &          & Primary  &            & Secondary   &           & Primary     &             & Secondary \\
\hline
$L_i/(L_1+L_2)_B$      &          &  0.940   &            & 0.060       &           & 0.948       &             & 0.052   \\
$L_i/(L_1+L_2)_V$      &          &  0.896   &            & 0.104       &           & 0.930       &             & 0.070   \\
$L_i/(L_1+L_2)_R$      &          &  0.860   &            & 0.140       &           & 0.919       &             & 0.081   \\
$L_i/(L_1+L_2)_I$      &          &  0.819   &            & 0.181       &           & 0.906       &             & 0.094   \\
$L_i/(L_1+L_2)_{TESS}$ &          & 0.832    &            & 0.168       &           & 0.911       &             & 0.089   \\
\hline
$r_{pole}$             &          & 0.0927   &            & 0.0814      &           & 0.2445      &             & 0.1385  \\
$r_{side}$             &          & 0.0928   &            & 0.0814      &           & 0.2470      &             & 0.1392  \\
$r_{point}$            &          & 0.0929   &            & 0.0816      &           & 0.2498      &             & 0.1411  \\
$r_{back}$             &          & 0.0929   &            & 0.0816      &           & 0.2490      &             & 0.1408  \\
\hline
LD law                 &          &          &            & Logarithmic &           &             &         & Square root \\
\hline
LD coefficients        & X1       & Y1       & X2          & Y2         & X1        & Y1          & X2         & Y2       \\
\hline
B                      & 0.754092 & 0.339897 & 0.846441    & 0.053593   & -0.113707 & 0.756935    &  0.106426  & 0.801234 \\
V                      & 0.652402 & 0.301196 & 0.761352    & 0.153674   & -0.081223 & 0.619344    &  0.113234  & 0.675540 \\
R                      & 0.553402 & 0.269000 & 0.690147    & 0.187412   & -0.066800 & 0.521842    &  0.075691  & 0.610059 \\
I                      & 0.443505 & 0.224402 & 0.603333    & 0.201823   & -0.050963 & 0.413470    &  0.037968  & 0.530766 \\
TESS                   & 0.549000 & 0.430897 & 0.678050    & 0.354726   & -0.402237 & 0.908710    &	-0.166004  & 0.831675 \\
\hline
\end{tabular}
\end{table*}

\section{Absolute parameters and evolutionary status}
\label{secEvolution}

The absolute parameters of EK~Cep and HS~Her derived from our analysis are shown in Table \ref{tableAbsPars}. The errors reported here are estimated by assuming that the errors in the adopted primary temperatures correspond to half a spectral class. This results in an error of about 200 K for the primary of EK~Cep and about 1000 K for the primary of HS~Her (fairly close to the estimate given by \citealt{cakirli2007}). We constructed models with primary temperatures at the limits of the range indicated by these errors (and all other parameters initialized with the values from Table \ref{tableModels}) and optimized them for the best fit to the observations. The errors given in Table \ref{tableAbsPars} represent the range of absolute parameters obtained from these models.

Our absolute parameters are in fairly good agreement with the results of previous studies, tabulated in Tables \ref{tableParsEKCep} and \ref{tableParsHSHer} for comparison. 

At first glance, it would appear that despite being based on far richer data sets than any of the previous studies, our results have relatively poor accuracy. This is, however, the result of uncritical and overoptimistic uncertainty estimates in the cited works. All the absolute parameter estimates for EK~Cep prior to our work were based on the radial velocity measurements made by \citet{tomkin1983} and \citet{ebbi1966b}, neither of which report measurement errors. In their, and subsequent analyses, these observations were treated as errorless, giving rise to unreasonably accurate separation and mass ratio determinations. In addition, a common practice in publishing the results of modeling eclipsing binaries is to report the formal fitting errors, often without making an attempt to estimate the external uncertainties, such as those in the determination of the temperatures.

The situation is similar in the case of HS~Her. All the analyses prior to the work of \citet{cakirli2007} were based on the radial velocity measurements made by \citet{cesco45}, that were also reported without errors and treated as perfect in later use. In addition, \citet{cesco45} detected only the primary component, which makes the accuracy of absolute parameters in some of the cited studies clearly dubious. The radial velocities reported by \citet{cakirli2007} do come with measurement errors, but they still give only the formal fitting errors for the absolute parameters.

We therefore argue that this is the first study to report the absolute parameters of EK~Cep and HS~Her with reasonable and realistic accuracy.

\begin{figure*}
    \centering
    \includegraphics[width=\textwidth]{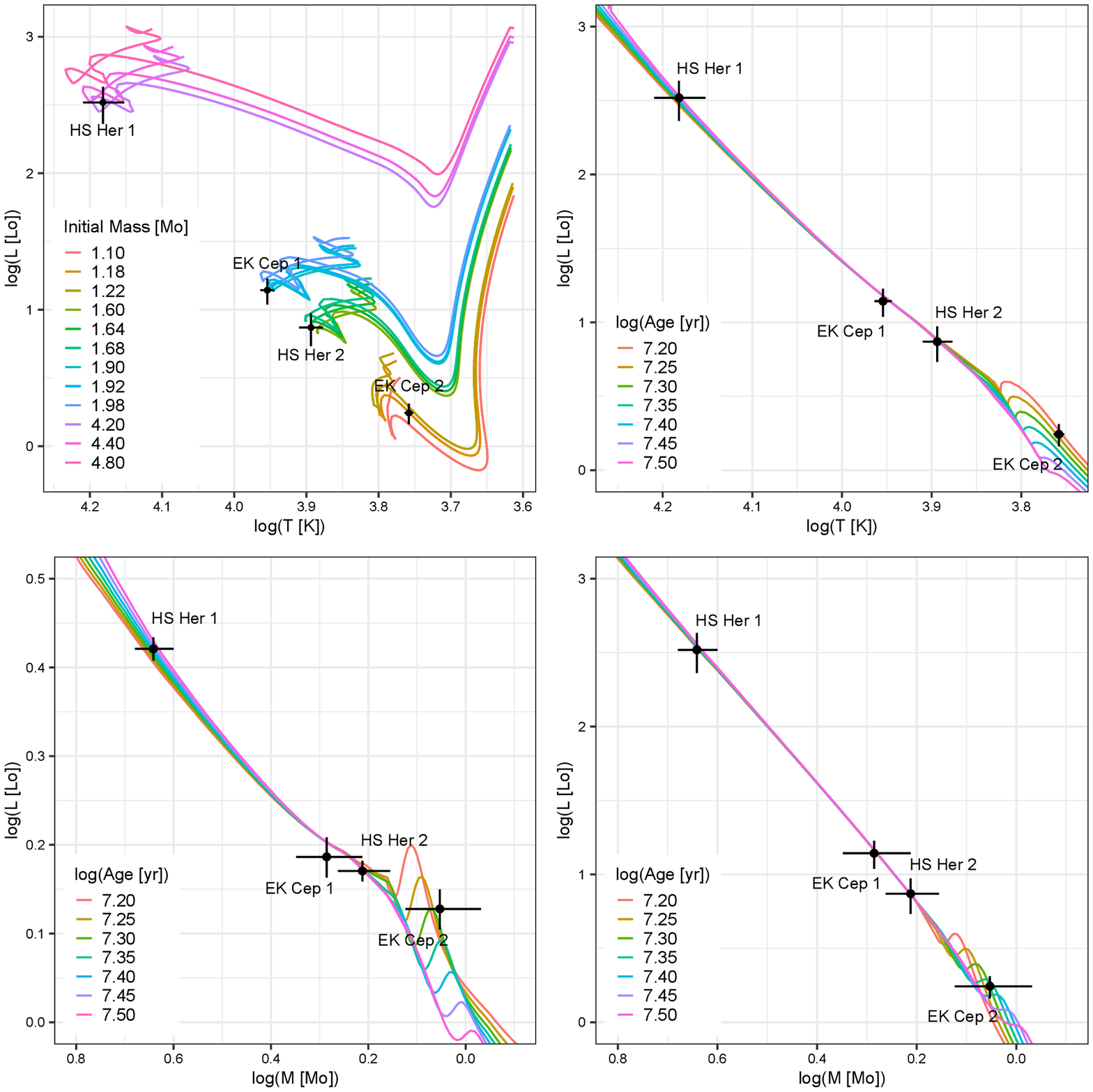}
    \caption{Top left: EK~Cep and HS~Her with the best-matching evolutionary tracks from the MIST library of models. Three closest tracks are shown for each component. Top right, bottom left and bottom right: EK~Cep and HS~Her with the best-matching MIST isochrones on the theoretical HR diagram, the mass-radius and mass-luminosity diagrams, respectively.}
    \label{figEvo}
    \end{figure*}

\begin{table}
\centering
\caption{Absolute parameters and distances of target stars.}
\label{tableAbsPars}
\begin{tabular}{lrrrrrr}
\hline
                    & EK~Cep &        & HS~Her   &       \\
\hline                                                     
\hline                                                     
Quantity            & Value  & Error  & Value    & Error \\
\hline                                                     
$M_1 [M_{\odot}]$   & 1.93   & 0.3    & 4.38     & 0.4   \\
$M_2 [M_{\odot}]$   & 1.13   & 0.2    & 1.63     & 0.2   \\
$R_1 [R_{\odot}]$   & 1.536  & 0.08   & 2.636    & 0.08  \\
$R_2 [R_{\odot}]$   & 1.342  & 0.07   & 1.481    & 0.04  \\
$L_1 [L_{\odot}]$   & 13.9   & 3      & 330      & 100   \\
$L_2 [L_{\odot}]$   & 1.75   & 0.3    & 7.4      & 2     \\
$T_1 [K]$           & 9000   & 200    & 15200    & 1000  \\
$T_2 [K]$           & 5730   & 80     & 7830     & 300   \\
$\log g_1$          & 4.35   & 0.3    & 4.24     & 0.2   \\
$\log g_2$          & 4.23   & 0.3    & 4.31     & 0.2   \\
$d [pc]$            & 164    & 17     & 498      & 75    \\
\hline
\end{tabular}
\end{table}

\begin{table*}
\centering
\caption{The absolute parameters of EK~Cep from past research and in this work.}
\label{tableParsEKCep}
\begin{footnotesize}
\begin{tabular}{lrrrrrrrrrr} 
\hline
\hline
Ref        
& $T_1 [K]$ 
& $T_2 [K]$ 
& $M_1 [M_{\odot}]$
& $M_2 [M_{\odot}]$
& $R_1 [R_{\odot}]$
& $R_2 [R_{\odot}]$
& $L_1 [L_{\odot}]$
& $L_2 [L_{\odot}]$
& $\log g_1 [cgs]$
& $\log g_2 [cgs]$ \\
\hline
$[1]$
& 9610(140)  & 5830(30) & 3.1(4.1) & 1.5(1.6) & 1.6(0.7)
& 1.3(0.6) & 20.42(3) & 1.9(3) & - & - \\
$[2]$ 
& - & - & 2.03(0.02) & 1.12(0.01) & 1.31(0.07) & 1.08(0.05) & - & - & - & - \\
$[3]$
& 9100 & 5800 & 2.03 & 1.12 & 1.58 & 1.31 & 15 & 1.5 & 4.35 & 4.25 \\
$[4]$
& - & - & 2.029(0.023) & 1.124(0.012) & 1.579(0.007) & 1.315(0.006)
& - & - & 4.349(0.01) & 4.251(0.006) \\
$[5]$
& 9000(200) & 5730(80) & 1.93(0.3) & 1.13(0.2) & 1.536(0.08) 
& 1.342(0.07) & 13.9(3) & 1.75(0.3) & 4.35(0.3) & 4.32(0.3) \\
\hline
\multicolumn{11}{l}{$[1]$ \cite{mezzetti1980}; $[2]$ \cite{tomkin1983}; $[3]$ \cite{hill1984}; 
$[4]$ \cite{andersen1991}; $[5]$ this paper.}
\end{tabular}
\end{footnotesize}
\end{table*}

\begin{table*}
\centering
\caption{The absolute parameters of HS~Her from past research and in this work.}
\label{tableParsHSHer}
\begin{footnotesize}
\begin{tabular}{lrrrrrrrrrr} 
\hline
\hline
Ref        
& $T_1 [K]$ 
& $T_2 [K]$ 
& $M_1 [M_{\odot}]$
& $M_2 [M_{\odot}]$
& $R_1 [R_{\odot}]$
& $R_2 [R_{\odot}]$
& $L_1 [L_{\odot}]$
& $L_2 [L_{\odot}]$
& $\log g_1 [cgs]$
& $\log g_2 [cgs]$ \\
\hline
$[1]$
& 16400 & - & - & 4.7(0.5) & 1.6 & 2.8 & 1.6 & - & - & - \\
$[2]$
& 15160 & 7560 & 6.5 & 1.9 & 3 & 1.7 & 427 & 8.9 & - & - \\
$[3]$
& 15200 & 7594(99) & 6(0.5) & 1.8(0.2) & 3.1(0.2) & 1.7(0.1)
& 445(60) & 10.4(0.6) & 4.25(0.15) & 4.23(0.15) \\
$[4]$
& 15500(300) & 7800(200) & 5(0.4) & 1.61(0.07) & 2.79(0.06)
& 1.59(0.04) & 417(77) & 8.7(1.18) & 4.244(0.016) & 4.241(0.01) \\
$[5]$
& 15200(750) & 7600(400) & 4.49(0.16) & 1.75(0.09) & 2.83(0.04)
& 1.61(0.02) & 386(1) & 8(1) & 4.19(0.01)& 4.27(0.01) \\
$[6]$
& 15200(1000) & 7830(300) & 4.38(0.4) & 1.36(0.2) & 2.636(0.08)
& 1.481(0.04) & 330(100) & 7.4(2) & 4.24(0.2) & 4.31(0.2) \\
\hline
\multicolumn{11}{l}{
$[1]$ \cite{hall71}; 
$[2]$ \cite{giuricin81}; 
$[3]$ \cite{bozkurt2006};
$[4]$ \cite{kali2006}} \\
\multicolumn{11}{l}{             
$[5]$ \cite{cakirli2007};
$[6]$ this paper. }
\end{tabular}
\end{footnotesize}
\end{table*}    






Compared to the main sequence calibrations for binary stars from \citet{eker2018}, the components of our binaries have somewhat atypical properties. For the primary of EK~Cep, the mass of $M_1 = 1.93 M_{\odot}$ corresponds to a main sequence star with a temperature slightly lower than the value we adopted (8150 vs 9000 K), larger radius (2.11 vs 1.54 $R_{\odot}$) and slightly higher luminosity (8.5 vs 7.4 $L_{\odot}$), a finding remarked upon previously by \citet{tomkin1983} and \citet{yildiz2003}. However, the secondary, at $M_2 = 1.13 M_{\odot}$, is expected to have a higher temperature (5960 vs 5730 K), smaller radius (1.24 vs 1.34 $R_{\odot}$), resulting in about the same luminosity (1.74 vs 1.75 $L_{\odot}$).

We examine the evolutionary status of our stars further using the MIST\footnote{MESA Isochrones \& Stellar Tracks, \textit{http://waps.cfa.harvard.edu/MIST/}} library of stellar evolution models \citep{mist2016a, mist2016b, mist2015}. As there are no indications otherwise, we assume solar metallicity for both systems and use the tabulations for rotating models.

In Fig. \ref{figEvo}, the components of EK~Cep and HS~Her are shown together with a selection of MIST evolutionary tracks for different initial stellar masses. For EK~Cep, the tracks closest to its components are $M_{1,i}\approx 1.92 M_{\odot}$ and $M_{2,i}\approx 1.18 M_{\odot}$; for HS~Her, $M_{1,i}\approx 4.4 M_{\odot}$ and $M_{2,i}\approx 1.64 M_{\odot}$. These initial masses are well within the errors of the current mass estimates from our binary system models, which is to be expected for very young stars. The primary of HS~Her has evolved somewhat off the ZAMS; the secondary of HS~Her and the primary of EK~Cep lie exactly on the ZAMS. The secondary of EK~Cep is still on a PMS track.

Fig. \ref{figEvo} also shows our stars together with the MIST isochrones for MS and PMS evolution phases in the HR diagram, the mass-radius and mass-luminosity planes. While the primary component of EK~Cep doesn't provide a reasonable constraint on the age of the system, the secondary lies on the isochrone with log(Age [yr])=7.2-7.3 (16-20 Myr). The components of HS~Her are located on the isochrone with log(Age [yr])=7.4-7.5 (25-32 Myr). These ages are in fair agreement with previous estimates.

Finally, we estimate the distances from our absolute parameters and the V magnitudes measured by \citet{hog2000} for EK Cep and \citet{oja1991} for HS Her. The bolometric correction is taken from \citet{eker2018}, and the extinction is interpolated from the Bayestar19 dust map \citep{green2019}. The derived distances are in good agreement with the Gaia parallaxes given in Table~\ref{tableGaia}: $164 \pm 17$ pc ($d_{Gaia} = 172 \pm 1$ pc) for EK Cep, and $498 \pm 75$ pc ($d_{Gaia} = 492 \pm 5$ pc) for HS Her.

\section{Conclusions}

Well-studied detached binaries in early stages of binary evolution and orbit circularization are valuable test cases for the theories of stellar structure and evolution. We conducted a comprehensive study of two young eclipsing binaries, EK~Cep and HS~Her, based on new photometric and spectroscopic observations and over half a century of eclipse timings. Both stars were studied in the past, and EK~Cep is known as the first (and for a long time the only) eclipsing binary with a PMS component. However, this is the first work to present high-quality multicolour CCD light curves and a modern analysis based on the Roche model of either object. 

The combined light curve and radial velocity curve fitting yields reliable orbital and stellar parameters of the binaries which are in excellent agreement with the results of an independent eclipse timing analysis. Both stars are slightly eccentric (e = 0.109 and 0.021 for EK~Cep and HS~Her, respectively), with obvious signs of apsidal motion. A comparison with evolutionary tracks shows that HS~Her is between 25 and 32 Myr old; its primary, B5 component, has already evolved somewhat off the ZAMS, but the A6.5 secondary lies exactly on the ZAMS. EK~Cep is between 16 and 20 Myr old; its primary is an A1.5 star on the ZAMS, and its secondary is a PMS star still contracting towards the main sequence. 

In addition to these results, we were able to rule out the hypothesized presence of a tertiary component in HS~Her based on updated eclipse timings and in the light of lacking spectroscopic evidence. This demonstrates the importance of monitoring close binaries with known eclipse timing variations and/or period variability. Continued observations of our targets in the following decades will allow us to gather eclipse timing data covering a full cycle of the apsidal motion for HS~Her, and put stronger constraints on the variation of orbital parameters in EK~Cep.

\section*{Acknowledgements}

The authors wish to thank S. Zervakos for his valuable help in data processing of TESS observations. We would also like to thank the anonymous referee, whose suggestions resulted in significant improvements of the paper. O.L. and A.C. acknowledge funding by the Ministry of Education, Science and Technological Development of Republic of Serbia (contract No. 451-03-68/2022-14/200002). H.M. acknowledges funding by the Bulgarian National Science Fund under contract DN 18/13-12.12.2017. Funding for the TESS mission is provided by NASA’s Science Mission directorate. This work has made frequent use of the Simbad database ({\it http://simbad.u-strasbg.fr/simbad/}), operated at the CDS, Strasbourg, France, and NASA's Astrophysics Data System Bibliographic Services ({\it http://adsabs.harvard.edu/}).

\section*{Data Availability}

The ground-based light curves and radial velocity curves of EK~Cep and HS~Her used in this investigation are available for download as machine-readable files alongside the paper on the journal website. The spectra can be made available on reasonable request.

\bibliographystyle{mnras}
\bibliography{hsher-ekcep}

\bsp	
\label{lastpage}
\end{document}